# Electronic excited states of protonated aromatic molecules: protonated Fluorene

Ivan Alata[1,3], Michel Broquier[1,2], Claude Dedonder[1,2], Christophe Jouvet[1,2] *, Ernesto Marceca[4]

1) CLUPS (Centre Laser de l'Université Paris Sud / LUMAT FR 2764) Bât. 106, Univ Paris-Sud 11 -91405 Orsay Cedex, France

2) Institut des Sciences Moléculaires d'Orsay (ISMO, UMR8624 CNRS )
 Bât. 210, Univ Paris-Sud 11 - 91405 Orsay Cedex, France

3) Atomic Energy Commission of Syria, Damascus, P.O. Box 6091 Syria. http://www.aec.org.sy/

4) INQUIMAE-FCEN, UBA, Ciudad Universitaria, 3er piso, Pab. II, (1428) Buenos Aires, Argentina

## Abstract

The photo-fragmentation spectrum of protonated fluorene has been recorded in the visible spectral region, evidencing an absorption that appears largely red shifted in comparison to that of the neutral molecule fluorene. The spectrum shows two different vibrational progressions, separated by 0.19 eV. As in the case of protonated linear polycyclic aromatic hydrocarbons (PAHs), comparison of the measured spectra with ab-initio calculations allows to associate the observed absorption shift with the charge transfer character of the excited state. The spectra can be properly simulated by geometry optimization of the ground and excited states, followed by Franck Condon analysis. The two vibrational bands progressions observed are assigned, with relatively good confidence, to the existence of two different conformers.

---

* Corresponding author: christophe.jouvet@u-psud.fr

# Introduction

The excited states of protonated aromatic molecules, in particular those of polycyclic aromatic hydrocarbons (PAHs) have been extensively studied in the condensed phase (1-7). However, the optical properties of protonated aromatic molecules isolated in the gas phase have been less studied experimentally, because of difficulties encountered in producing sufficient amount of these reactive species (8-10). From another point of view, recent developments in quantum calculation methods have improved the description of the excited states of such species (11-15) and comparison of these calculations with gas phase spectroscopic data is now possible.

The photophysics of protonated polycyclic aromatic hydrocarbons became attractive since it was suggested that these ions might be present in the interstellar medium (14,16,17) being potential carriers for the diffuse interstellar bands (DIBs) that appear in astronomical observations.

We have recently studied the electronic spectra of linear protonated PAHs for which the $S_1$-$S_0$ transitions are all in the visible region and they do not show a monotonic red shift as a function of the molecular size, in contrast to their neutral analogues (18). This behaviour was interpreted using the results of ab-initio calculations as due to the charge transfer character found in the excited states of PAHs, especially those with an even number of aromatic rings.

Previous work was focussed in protonated PAHs composed by different number of condensed six-membered rings. In the present paper, we will investigate the behaviour of protonated fluorene, a PAH analogue having a five-membered ring in the structure, with particular interest in the effect of this moiety on the protonation site and on the charge transfer character of the excited state.

Protonation of fluorene has been used to change the optical properties of conjugated polymers (19). This allows the design and development of new organic emitter and sensor materials tailored toward particular needs. The effect of protonation on the monomer might be useful to understand the optical properties of larger systems.

Finally, many efforts have been done recently to characterise the stability of the different conformers of protonated molecules through their infrared vibrational spectroscopy (20-24). We will show in the present paper that the recent progress achieved by ab-initio methods to describe an excited state is such that the electronic spectroscopy nowadays represents a quite accurate alternative to assign conformers, as long as the excited states are not too short lived, i.e. some vibrational analysis can be performed.

## Methodology

### Experimental details

Protonated fluorene is produced in a pulsed high voltage electrical discharge source coupled to a pulsed nozzle (General valve) of 300 μm in diameter. The discharge is produced between two electrodes positioned 3 mm downstream from the nozzle. The gas mixture consists of 50% of Helium and 50% of $H_2$ seeded with fluorene vapor, obtained by heating the solid at about 120° C. The protonated species are obtained only in presence of $H_2$. The typical backing pressure in the source is about 4 bars, while under operation the pressure in the source chamber is $1\times10^{-5}$ mbar. The expanding molecular beam containing the ions passes through a 1 mm diameter skimmer located 10 cm downstream from the nozzle orifice, and the collimated ion beam passes into a second differentially pumped chamber ($p = 2\times10^{-6}$ mbar).

The ions produced in the source are extracted with pulsed voltages (2 kV) into a reflectron time of flight mass spectrometer, perpendicularly to the molecular beam axis. A reasonably good mass resolution ($m/\Delta m > 400$) is achieved at the entrance of the reflectron mirror, after the ions have flown a distance of 157 cm along the field-free flight tube. At this point, the protonated species are well separated from the corresponding radical cations ($\Delta m/Z = 1$ between them), which are also formed in the discharge.

Photo-fragmentation spectra of the protonated species are recorded by monitoring the intensity of neutral fragments originated during the interaction with a tuneable laser, as a function of its wavelength. This process is carried out inside a field-free 2 cm long interaction region, located in front of the reflectron plates (see the Electronic Supplementary Information for a scheme of the experimental set-up). The fragmentation process is done using a 10 Hz, nanosecond OPO laser (Euroscan) pumped with the third harmonic (355 nm) of a $Nd^{+3}$:YAG laser (Quantel YG981C). The OPO laser delivers a tuneable wavelength range from 413 to 670 nm with an output power in the range of 1 to 15 mJ. The laser is mildly focused to intersect the ion bunch in a volume of typically 1 $mm^3$. In these conditions the power dependence of the signal evidenced a small saturation, but lower laser energies proved being insufficient to attain the intensity demanded by the experiment.

The detection of neutral fragments is carried out on a MCP detector located directly behind the reflector mirror. The signal from the MCP detector is sent to the digitizing storage oscilloscope interfaced to a PC.

Photo-fragmentation of mass selected ions is not the only source of neutrals arriving at the detector. Neutral fragments may also be generated by collisions with the residual gas during the time of flight. To discriminate the neutral fragments produced by collisions from those produced by laser absorption, the parent ions are post-accelerated to a final energy of 5 kV immediately before they enter the interaction region. In this way, the kinetic energy of the neutral fragments produced by laser absorption will be 5 kV, while the fragments produced by collisions in the flight tube will be sensibly slower (2 kV), and hence it becomes easy to differentiate both signals. The interaction region is short enough to assume that the number of neutrals produced by collisions in this region can be neglected.

The signal to noise ratio of the experiment is low and therefore it is necessary to average many scans in order to get a spectrum like that presented in figure 1.

## Calculations

*Ab initio* calculations have been performed with the TURBOMOLE program package (25), making use of the resolution-of-the-identity (RI) approximation for the evaluation of the electron-repulsion integrals (26). The equilibrium geometry of the ground electronic state ($S_0$) has been determined at the MP2 (Moller-Plesset second order perturbation theory) level. Excitation energies and equilibrium geometry of the lowest excited singlet state ($S_1$) have been determined using a second-order approximate coupled-cluster calculation (RI-CC2). The CC2 method was chosen because it gives reasonable results for medium size closed shell molecules for a moderate computational time (27). Besides it allows excited state geometry optimization more easily than the CASPT2 method. Calculations were performed with the SVP basis set (7s4p1d for carbon and 4s1p for hydrogen), which is quite small but gives reasonable good results for this type of systems (18). Some complementary calculations have been performed with the aug-cc-pVDZ basis set (28). The isomers are noted $FC_nH^+$, depending on the carbon atom where the proton is attached to, the numbering of the carbon atoms being shown in the insert of figure 1.

For the two most stable isomers, a complete vibrational analysis has been performed after the excited-state optimization, the vibrational frequencies were calculated in both the ground and the excited states, and a Franck-Condon analysis was performed using the Pgopher software, a program for simulating and fitting rotational, vibrational and electronic spectra (29).

# Results

## Experimental

The experimental spectrum is presented in figure 1: the laser has been scanned to the red and no other bands were detected. Clearly, the spectrum shows two vibrational progressions: a band system A, starting at 479.8 nm (2.58 eV), and a more intense band system B, starting at 447.6 nm (2.77 eV). The vibrational progression found in system B appears well structured, whereas the progression in system A hardly gets out of the experimental noise, although it can be noted a clear onset with a doublet band. The vibronic spectrum is superimposed on a broad background signal, the existence of which will be discussed in the next section.

Using a simple thermodynamic calculation involving tabulated values for the ionization potential (30) and proton affinity (31) of fluorene, it was possible to estimate a value of 2.93 eV for the dissociation threshold of protonated fluorene in fluorene$^{+\bullet}$ + H$^{\bullet}$ (assuming this dissociation channel is major as for other protonated polycyclic aromatic molecules (32,33)), which lie above the position of the observed first vibronic bands, meaning that the recorded photo-fragmentation spectrum results from a resonance enhanced multi-photon process. This is also evidenced by the marked dependence of the signal on the laser power, as mentioned before in the experimental section, changing from being mildly saturated to a total absence of signal for low laser powers.

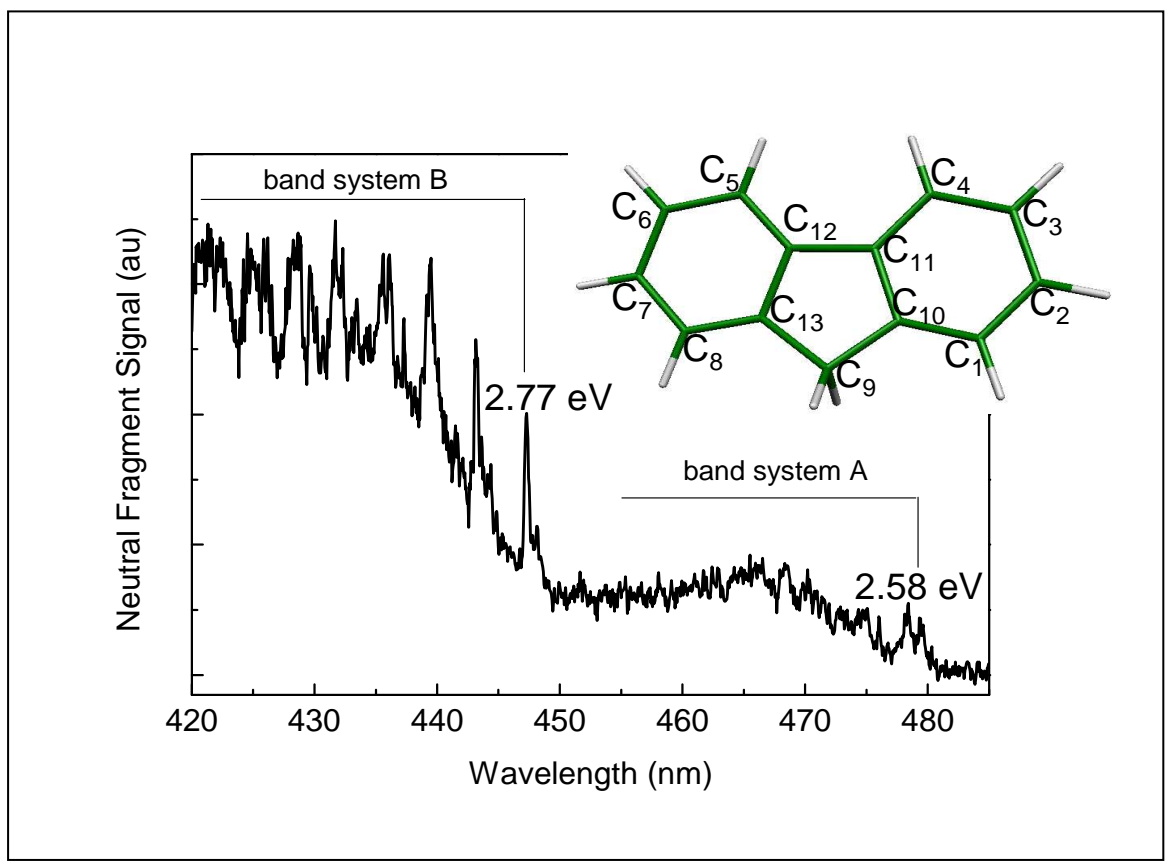

Figure 1: Photo-fragmentation spectrum of protonated fluorene. The numbering of the carbon atoms is represented on the neutral fluorene molecule in the inset.

As for linear protonated PAHs, the electronic absorption is strongly red shifted as compared to the absorption of neutral fluorene (296 nm, 4.18 eV) (34,35) the magnitude of the shift being quite similar to the one observed in the case of Naphthalene (18).

The band systems A and B found for protonated fluorene may correspond either to excitations to different electronic states or to the presence of different isomers. We will make use of ab-initio calculations and comparison between recorded and simulated spectra to identify which of the two hypotheses gives the best agreement.

## Ab initio calculations

The results of the ab-initio calculations, in particular the ground state relative stabilities and transition energies, are presented in table 1 for different isomers of protonated fluorene. The ground state energies of the different isomers are given relative to the most stable $FC_2H^+$ isomer. There are two isomers with similar ground state energies, $FC_2H^+$ and $FC_4H^+$, while the other isomers have much higher ground state energies, in agreement with previous calculations (36) of proton affinities of fluorene in the ground state. The stability of isomers $FC_2H^+$ and $FC_4H^+$ can be understood in terms of *para* and *ortho* conjugation effect across the planar biphenyl framework relative to carbon $C_{11}$ (and $C_{12}$). Isomers $FC_2H^+$ and $FC_4H^+$ have the highest calculated $S_1 \leftarrow S_0$ transitions energies, which are close to the observed band origins. A rough trend is observed when considering the stability and the $S_1 \leftarrow S_0$ transition energies of the isomers, the less stable isomers having the lowest transition energies. This has also been observed in the case of linear PAHs (18) and might be a quite general behaviour, the origin of which must be investigated in more detail.

*Table 1: Ab-initio calculations for the different isomers of protonated fluorene (see the insert of figure 1 for the numbering of the carbon atoms). The ground state energies are calculated at the MP2 level and referenced to the most stable $FC_2H^+$ isomer. Vertical $S_1 \leftarrow S_0$ and $S_2 \leftarrow S_0$ transition energies as well as adiabatic $S_1 \leftarrow S_0$ transition energies are calculated at the CC2 level with the SVP basis set. The values in parenthesis are the oscillator strengths of the transitions. The difference between the zero point energies of the ground and the excited states is used to correct the calculated transition energy, in order to compare the calculated value with the experimental measurements.*

|  | Ground state relative energy | $S_1 \leftarrow S_0$ Vertical transition | $S_2 \leftarrow S_0$ Vertical transition | $S_1 \leftarrow S_0$ Adiabatic transition | Zero point energy difference δZPE | $S_1 \leftarrow S_0$ Adiabatic transition corrected for δZPE | $S_1 \leftarrow S_0$ Adiabatic transition corrected for δZPE |
|---|---|---|---|---|---|---|---|
|  | (eV) | (eV) | (eV) | (eV) | (eV) | (eV) | (nm) |
| $FC_2H^+$ | 0.00 | 3.20 | 3.5 | 2.91 (.029) | -0.156 | 2.76 | 449.2 |
| $FC_4H^+$ | 0.09 | 2.91 | 3.08 | 2.74 (.046) | -0.155 | 2.59 | 478.7 |
| $FC_{10}H^+$ | 0.31 | 2.91 | 3.20 | 2.65 (.196) | -0.157 | 2.49 | 497.9 |
| $FC_3H^+$ | 0.40 | 2.46 | 2.76 | 2.22 (.158) | -0.101 | 2.12 | 584.8 |
| $FC_1H^+$ | 0.49 | 2.33 | 2.59 | 2.07 (.059) | -0.103 | 1.97 | 629.4 |
| $FC_{11}H^+$ | 0.83 | 2.81 | 3.09 | 2.10 (.001) | -0.093 | 2.01 | 616.8 |

The optimized geometry of the $FC_1H^+$, $FC_2H^+$, $FC_1H^+$ and $FC_4H^+$ isomers in both the ground and the excited states maintains the plane of the carbon atoms as a symmetry plane, while the $FC_{10}H^+$ and $FC_{11}H^+$ isomers have a very distorted geometry with a large geometry change between the ground and excited state. The ground and excited state geometries of the two most stable isomers, $FC_2H^+$ and $FC_4H^+$, are given in the supplementary information.

Since ground state and transition energies of isomers $FC_2H^+$ and $FC_4H^+$ are very close, complementary calculations have been performed with the aug-cc-pVDZ basis set, to see if there is a strong basis set dependence for the ordering of the transitions and the energy gap.

Table 2 shows that the ground state energy difference between the two isomers stays the same, and the transition energies get lower with the size of the basis set. In spite of that, the ordering of the transitions does not change, and the energy gap between the two transitions remains close to the experimental gap between the two band systems (2.77-2.58 = 0.19 eV).

*Table 2: Comparison of the ground and first excited state energies for the isomers $FC_2H^+$ and $FC_4H^+$ of protonated fluorene using two different basis sets. The energies are given in eV.*

|  | SVP basis set | | | | aug-cc-pVDZ basis set | | | |
| --- | --- | --- | --- | --- | --- | --- | --- | --- |
|  | Ground state | $S_1 \leftarrow S_0$ Vertical | $S_1 \leftarrow S_0$ Adiabatic | Difference in transition energy | Ground state | $S_1 \leftarrow S_0$ Vertical | $S_1 \leftarrow S_0$ Adiabatic | Difference in transition energy |
| $FC_2H^+$ | 0 | 3.08 | 2.91 | 0.18 | 0 | 3.12 | 2.84 | 0.20 |
| $FC_4H^+$ | 0.095 | 2.91 | 2.73 |  | 0.093 | 2.82 | 2.64 |  |

In addition, the successive isomerization barriers between isomers $FC_2H^+$ and $FC_3H^+$ and between $FC_3H^+$ and $FC_4H+$ have been estimated at the B3-LYP/SVP level of theory. These barriers are about 1 eV, roughly 3 times higher than in protonated benzene (37)(the potential energy profile is presented in the electronic supplementary material).

## Discussion

In all the spectra recorded with our set-up, the structured vibrational progressions are superimposed on a continuous background with a comparable intensity. We ascribe this effect to the presence of protonated molecules formed under very different cooling conditions, the colder ions being responsible for the vibrational band structure while the hot ones contribute to the background. The wide spread in the internal energy of protonated molecules produced by the electric discharge in a supersonic expansion has to be explored further to characterize the ions sampled in the experiment (a qualitative model is presented in the electronic supplementary material).

From the calculations, there are two possibilities for the assignment of the observed two band systems A and B: either they correspond to transitions of two different isomers of protonated fluorene or to two different excited states of a single isomer.

We may first assume that two isomers are responsible for the observed spectrum. A first argument used to identify the isomers in the ion source is based on their calculated ground state energy, which should not be too high as compared to the most stable isomer. According to this criterion, $FC_2H^+$ and $FC_4H^+$ isomers are then expected to be observed in the experiment since their ground state energies only differ in 0.095 eV, for the other isomers the difference is larger than 0.3 eV. This is in line with previous studies on protonated molecules (18), where only the lowest energy isomers have been observed. The second argument takes into account the calculated adiabatic transition energies, which should be close to the experimental origins of the band systems, when the difference in the zero point energy ($\delta$ZPE) between $S_0$ and $S_1$ is taken into account. The transition for the most stable $FC_2H^+$ isomer is calculated at 2.76 eV, which almost coincides with the band origin of system B, the most intense progression, observed at 2.77 eV. There is also a very good agreement in the case of the $FC_4H^+$ isomer, for which the calculated transition energy is 2.59 eV and the origin of band system A is observed at 2.58 eV. On the other hand, the values of the transition energies calculated for the isomers $FC_1H^+$, $FC_3H^+$ and $FC_{11}H^+$ are much lower than the experimental band origins and therefore we consider that these isomers do not contribute to the observed spectrum. Although the transition energy of the $H_{10}^+$ isomer is closer to the first experimental band origin, its presence is unlikely because its ground state energy is 0.31 eV higher than that of the most stable $FC_2H^+$ isomer.

The ground state relative energies and the transitions energies have also been calculated with the aug-cc-pVDZ basis set for the $FC_2H^+$ and $FC_4H^+$ isomers to test the relative stability of the isomers and the ordering of the excited state transition energies. As shown in table 2, the $FC_2H^+$ isomer remains the most stable species and the ground state energy difference between $FC_4H^+$ and $FC_2H^+$ changes only 0.002 eV. The transitions energies of these two isomers are lower with this larger basis set, but the $FC_4H^+$ transition remains to the red of the $FC_2H^+$ transition, and the gap between them (0.2 eV) stays close to the experimental value.

The other hypothesis is to assign the two band systems to two different excited states, i.e. $S_1$ and $S_2$, of the most stable isomer. This possibility was explored by performing the $S_1$ and $S_2$ excited state optimizations. Table 3 summarizes the calculated adiabatic transition energies, 2.76 eV for $S_1$ and 3.26 eV for $S_2$ when the states $S_1$ and $S_2$ are optimized independently and when corrections for the difference in the zero point energies are applied. In addition, the $S_1$-$S_2$ energy gap, which is 0.3 eV for the vertical transition, becomes even larger (0.50 eV) when the excited states are optimized, far more than the observed gap (0.19 eV).

*Table 3: Excited state transition energies for the $FC_2H^+$ isomer. The values in bold italic have been corrected for the zero point energy difference. All values are given in eV.*

|  | $S_0$ | $S_1$ | $S_2$ | $S_1 \leftarrow S_0$ Adiabatic transition | $S_2 \leftarrow S_0$ Adiabatic transition | $S_2 \leftarrow S_1$ Vertical energy gap | $S_2 \leftarrow S_1$ Adiabatic energy gap |
|---|---|---|---|---|---|---|---|
| $S_0$ optimized geometry | 0 | 3.20 | 3.50 |  |  | 0.30 |  |
| $S_1$ optimized geometry | 0.27 | 2.64 | 3.4 | 2.91 ***2.76*** |  |  | 0.47 ***0.50*** |
| $S_2$ optimized geometry | 0.13 | 3.08 | 3.25 |  | 3.38 ***3.26*** |  |  |

The previous criteria are linked to the energetic of the system. We can also use the change in geometry between the ground and the excited states to further test if our assignment is reasonable. This means that the simulated spectra, obtained via the Franck-Condon (FC) factors calculated using the ground and excited state frequencies, should resemble the experimental traces, either if different isomers are present or if different excited states are involved.

Figure 2a shows the simulated spectra obtained for the $S_1$-$S_0$ transition of the $FC_2H^+$ isomer by convolution of the stick spectrum with a Gaussian profile of 10 cm$^{-1}$ width, at temperatures of 0 and 300 K compared with the most intense observed progression, band system B,. The agreement between the experimental data and the theoretical calculation is quite good, and this strongly supports the assignment of band system B to the most stable $FC_2H^+$ isomer in the first excited state.

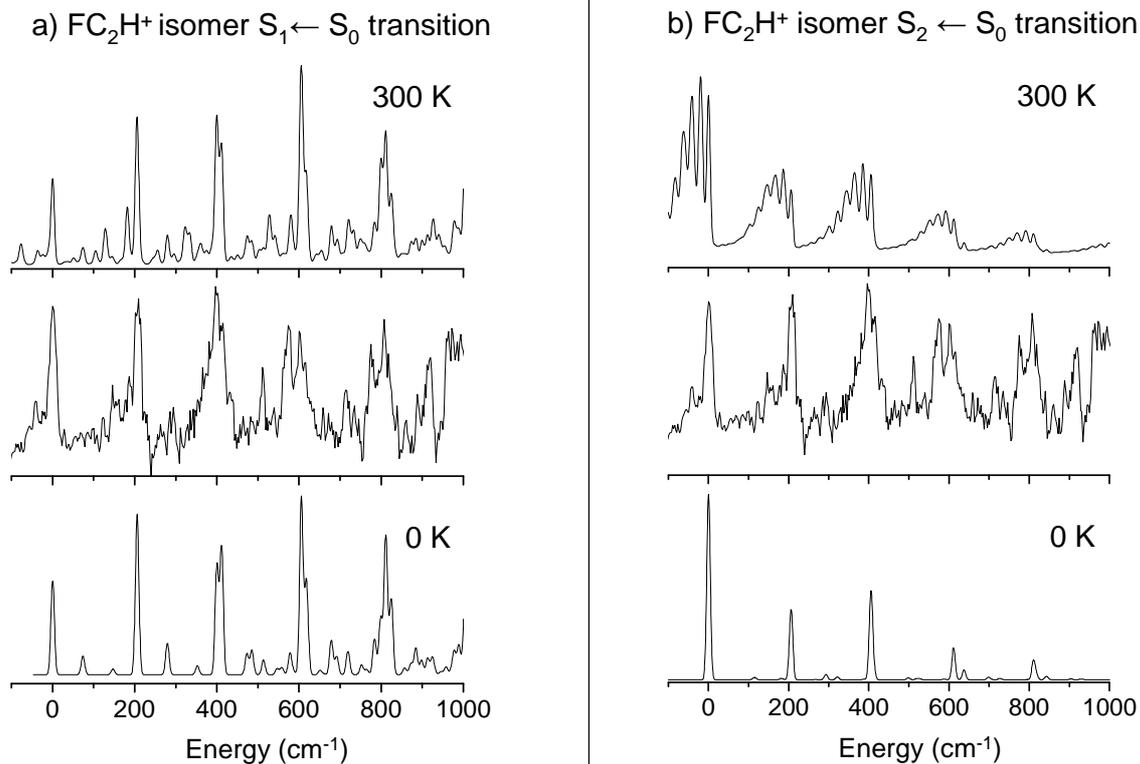

Figure 2

Figure 2: Franck Condon simulation of the vibronic spectrum. a) Comparison of the experimental band system B (middle trace, background subtracted) with the simulated spectrum corresponding to the $S_1 \leftarrow S_0$ transition of the $FC_2H^+$ isomer at 0 K (lower trace) and 300 K (upper trace). b) Comparison between the experimental band system B (middle trace, background subtracted) and the simulated spectrum corresponding to the $S_2 \leftarrow S_0$ transition of the $FC_2H^+$ isomer at 0 K (lower trace) and 300 K (upper trace).

On the other hand, a poorer agreement is found when the experimental band system B is compared with the simulated spectrum obtained for the $S_2 \leftarrow S_0$ transition of the $FC_2H^+$ isomer (see figure 2b). Among the differences are: i) band intensities decrease with the energy only in the simulated spectrum, and ii) a sequence of hot bands (-20 cm$^{-1}$) appears more markedly in the simulated spectrum when the temperature is increased. These bands correspond to the lowest out of plane vibrations of 76 and 113 cm$^{-1}$ in $S_0$, and 57 and 91 cm$^{-1}$ in $S_2$), for which the Franck Condon factors are greater in the $S_2 \leftarrow S_0$ transition than in the $S_1 \leftarrow S_0$ transition because the geometry change is smaller between $S_0$ and $S_2$ than between $S_0$ and $S_1$. On the whole, the agreement between the observed progression and the simulated spectra is more satisfactory if band system B is assigned to the $S_1$ state rather than to the $S_2$ state of the $FC_2H^+$ isomer.

The simulated spectrum for the $FC_4H^+$ isomer has also been calculated, The main difference with the $FC_2H^+$ isomer being the width of the bands which are broadened by hot bands. Since band system A is weakly observed and hardly resolved, the comparison between the simulated spectra for the $FC_2H^+$ and $FC_4H^+$ isomers is not very demonstrative, although the first doublet, where the first band is a hot band, seems to be best reproduced by the $FC_4H^+$ simulated spectrum (see figure SI-4 in the Electronic Supplementary Information).

Thus, the Franck Condon simulations do not contradict the assignment of the stronger band system B to the $FC_2H^+$ isomer and the weaker band system A to the $FC_4H^+$ isomer, its intensity being weaker because of the 0.09 eV higher ground state energy, but the energy arguments are more convincing than the Franck Condon simulations for the band system A.

It could also be questioned whether band system A could correspond to hot bands However, if we try to simulate the $FC_2H^+$ spectrum with a temperature high enough to have hot bands populated 1500 cm$^{-1}$ below the origin, then the region of the 0-0 transition is much more congested than what is observed.

The vibrational assignment of the bands in system B can be deduced from the calculated spectra for the $S_1 \leftarrow S_0$ transition of the $FC_2H^+$ isomer and is presented in table 4. The 2 most actives modes are shown in figure 3. They correspond to in plane symmetric deformations very similar to the modes that have been observed for protonated naphthalene and tetracene (18). Out of plane modes are only weakly active, like mode 66 and mode 63, and are responsible of the hot bands appearing as shoulders on the red side of the main bands. Mode 43 at 205 cm$^{-1}$ is also quite similar to the mode observed in the $S_1 \leftarrow S_0$ transition of the neutral molecule (208 cm$^{-1}$) (34) but is more active in the protonated species, which implies the occurrence of a stronger deformation upon excitation (the calculated ground and excited state frequencies are listed in the electronic supplementary information).

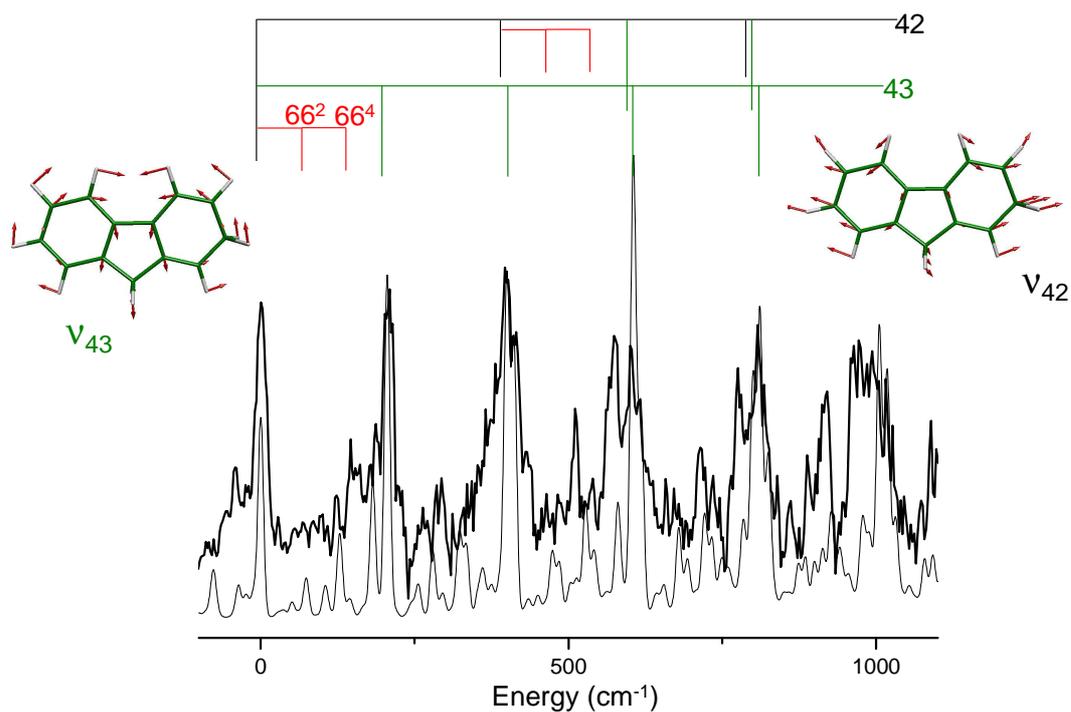

Figure 3

Figure 3: Assignment of the vibrational transitions: the upper trace is the experimental spectrum and the lower trace is the calculated one; on the top is shown the assignment of the vibrational transitions and the actives modes responsible for most of the vibrational progression are drawn in the insets.

*Table 4: assignment of the vibrational bands in band system B*

| Experimental frequency (cm$^{-1}$) | Calculated frequency (cm$^{-1}$) | Assignment |
|---|---|---|
| -39 | -30 | $63_1^1$ |
| 0 | 0 | $0_0^0$ |
| 209 | 206 | $43_0^1$ |
| 397 | 399 + 411 | $42_0^1 + 43_0^2$ |
| 511 | 485 | $66_0^2 43_0^2$ |
| 574 | 577 | $39_0^1$ |
| 602 | 606 + 616 + 619 | $43_0^1 42_0^1 + 43_0^3 + 38_0^1$ |
| 775 | 783 + 799 | $39_0^1 43_0^1 + 42_0^2$ |
| 807 | 810 + 822 | $43_0^2 42_0^1 + 43_0^4$ |
| 990 | 977 + 1004 + 1016 | $39_0^1 42_0^1 + 43_0^1 42_0^2 + 43_0^3 42_0^1$ |

The role of the charge transfer nature of the first excited state has been highlighted previously for protonated linear PAHs (18). The same is true in the case of protonated fluorene, as shown in figure 4, where the HOMO and LUMO orbitals of the $FC_2H^+$ isomer are depicted for comparison. It is observed that the $S_1 \leftarrow S_0$ transition mainly corresponds to a HOMO-LUMO excitation with an electron transfer from the non protonated benzene ring to the protonated one. This charge transfer is probably responsible for the large deformation occurring in the molecule upon excitation, which is more important in the protonated species than in the neutral molecule. The localisation of the electron on the non protonated part of the molecule has been explained in the case of the protonated benzene dimer (38) in terms of the larger ionisation potential of the protonated benzene moiety with respect to that of the neutral homologue, resulting from the presence of a positive charge in one of the rings. As a consequence, the HOMO is more stabilized in the protonated moiety and the first electronic transition basically corresponds to a transition from the neutral part HOMO towards the protonated part LUMO. This also explains why the electronic transitions are red shifted as compared to those in the neutral homologues.

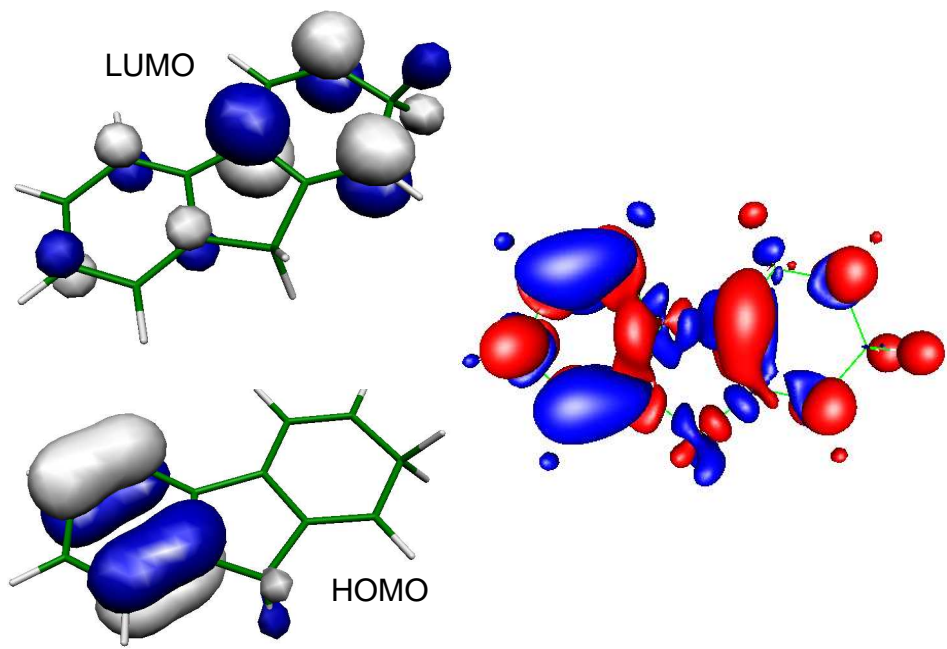

Figure 4

Figure 4: Orbitals involved in the first excited state transition of protonated fluorene. Left: Scheme of the Highest Occupied Molecular Orbital (HOMO) and Lowest Unoccupied Molecular Orbital (LUMO) involved in the $S_1 \leftarrow S_0$ transition for the $FC_2H^+$ isomer. Right: Change in the charge density between the ground and the excited states. The contour in red corresponds to an increase of 0.02 q (more negative charge) and in blue the opposite case. Clearly, the electronic density is increasing on the protonated benzene ring upon excitation.

The Diffuse Interstellar Bands (DIBs) observed in the visible to the near infrared spectral range (39) provide a spectroscopic source of information about the molecules present in the interstellar medium. However, none of them could be assigned to molecular carriers so far, and it has been suggested that protonated PAHs could be possible carriers for the DIBs (16,17). One of the consequences of the charge transfer character of the first excited state is that a relatively large geometry change is produced during the electronic excitation, resulting in an important vibrational activity and fairly low oscillator strengths. On the other hand, the DIBs spectra present no apparent regularity in the line spacing, which means that they cannot be originated in a molecular system with very active Franck Condon progressions. Thus, the first excited state of protonated fluorene is not likely to be responsible for any DIB (a comparison between the observed spectrum and the wavelengths of the DIBs (39) is presented in the supplementary information).

As for other PAHs, protonation leads to a strong red shift of the electronic absorption, which means that the proton affinity of the excited state is much larger than that of the ground state (8,9). The proton affinity of fluorene in the ground state is 8.62 eV (31) and becomes 10.03 eV in the excited state (the energy of the $S_1 \leftarrow S_0$ transition in the neutral is 4.18 eV (34,35) and it is 2.77 eV for the protonated molecule; PA*($S_1$)= 8.62 + (4.18 -2.77) =10.03 eV for the most stable isomer). The effect of protonation on the optical properties of conjugated fluorene–pyridine copolymers (19) is to induce a large red shift of the fluorescence emission from 3.2 eV for non protonated polymers, to 2.6 eV for protonated polymers. In the calculations (19) done to understand this effect, the proton is located on the pyridine moiety since the proton affinity of pyridine, 9.64 eV (31), is higher than the proton affinity of fluorene in its ground state, but lower than the proton affinity of fluorene in its excited state. Besides, the protonation of the pyridine molecule does not induce a shift of the electronic absorption, and the proton affinity of the excited state of pyridine should be similar to the ground state proton affinity. Thus, it is energetically favourable to locate the proton on the fluorene ring in the excited state, which implies that an excited state proton transfer is quite probable. This effect has not been considered previously (19) due the lack of knowledge on the optical properties of protonated molecules and it might be interesting to test this hypothesis by performing time resolved experiments.

## Conclusions

Protonation of fluorene leads to a strong red shift of the electronic transition as compared to the neutral molecule. Two vibrational band systems are observed and assigned by comparison with ab-initio calculations to the first excited state transitions of two different isomers. The agreement between the calculations (at the level of theory and basis set used in this paper) and the experiment is extremely good for the transition energies, as it has already been observed for linear protonated PAHs. The large red shift is due to the strong charge transfer character of the excited state as in other protonated PAHs. This shift of the electronic absorption implies a large increase of the excited state proton affinity, which can shade new perspectives on the excited state dynamics of protonated copolymers containing fluorene.

## Acknowledgments

The authors thank Prof. O. Dopfer for helpful discussion. This work has been supported by the Université Paris–Sud 11, by the ANR research grant (ANR2010BLANC040501), the PROCOPE 17832NK program and the RTRA "triangle de la physique". I.A. thanks the Atomic Energy Commission of Syria for financial support. The calculations have been performed on the GMPCS cluster of LUMAT. E.M. has been supported by the CONICET/CNRS exchange program.

Figure caption

Figure 1: Photo-fragmentation spectrum of protonated fluorene. The numbering of the carbon atoms is represented on the neutral fluorene molecule in the inset.

Figure 2: Franck Condon simulation of the vibronic spectrum. a) Comparison of the experimental band system B (middle trace, background subtracted) with the simulated spectrum corresponding to the $S_1 \leftarrow S_0$ transition of the $FC_2H^+$ isomer at 0 K (lower trace) and 300 K (upper trace). b) Comparison between the experimental band system B (middle trace, background subtracted) and the simulated spectrum corresponding to the $S_2 \leftarrow S_0$ transition of the $FC_2H^+$ isomer at 0 K (lower trace) and 300 K (upper trace).

Figure 3: Assignment of the vibrational transitions: the upper trace is the experimental spectrum and the lower trace is the calculated one; on the top is shown the assignment of the vibrational transitions and the actives modes responsible for most of the vibrational progression are drawn in the insets.

Figure 4: Orbitals involved in the first excited state transition of protonated fluorene. Left: Scheme of the Highest Occupied Molecular Orbital (HOMO) and Lowest Unoccupied Molecular Orbital (LUMO) involved in the $S_1 \leftarrow S_0$ transition for the $FC_2H^+$ isomer. Right: Change in the charge density between the ground and the excited states. The contour in red corresponds to an increase of 0.02 q (more negative charge) and in blue the opposite case. Clearly, the electronic density is increasing on the protonated benzene ring upon excitation.

Table 1: Ab-initio calculations for the different isomers of protonated fluorene (see the insert of figure 1 for the numbering of the carbon atoms). The ground state energies are calculated at the MP2 level and referenced to the most stable $FC_2H^+$ isomer. Vertical $S_1 \leftarrow S_0$ and $S_2 \leftarrow S_0$ transition energies as well as adiabatic $S_1 \leftarrow S_0$ transition energies are calculated at the CC2 level with the SVP basis set. The values in parenthesis are the oscillator strengths of the transitions. The difference between the zero point energies of the ground and the excited states is used to correct the calculated transition energy, in order to compare the calculated value with the experimental measurements.

|  | Ground state relative energy (eV) | $S_1 \leftarrow S_0$ Vertical transition (eV) | $S_2 \leftarrow S_0$ Vertical transition (eV) | $S_1 \leftarrow S_0$ Adiabatic transition (eV) | Zero point energy difference $\delta$ZPE (eV) | $S_1 \leftarrow S_0$ Adiabatic transition corrected for $\delta$ZPE (eV) | $S_1 \leftarrow S_0$ Adiabatic transition corrected for $\delta$ZPE (nm) |
|---|---|---|---|---|---|---|---|
| $FC_2H^+$ | 0.00 | 3.20 | 3.50 | 2.91 (.029) | -0.156 | 2.76 | 449.2 |
| $FC_4H^+$ | 0.09 | 2.91 | 3.08 | 2.74 (.046) | -0.155 | 2.59 | 478.7 |
| $FC_{10}H^+$ | 0.31 | 2.91 | 3.20 | 2.65 (.196) | -0.157 | 2.49 | 497.9 |
| $FC_3H^+$ | 0.40 | 2.46 | 2.76 | 2.22 (.158) | -0.101 | 2.12 | 584.8 |
| $FC_1H^+$ | 0.49 | 2.33 | 2.59 | 2.07 (.059) | -0.103 | 1.97 | 629.4 |
| $FC_{11}H^+$ | 0.83 | 2.81 | 3.09 | 2.10 (.001) | -0.093 | 2.01 | 616.8 |

Table 2: Comparison of the ground and first excited state energies for the isomers $FC_2H^+$ and $FC_4H^+$ of protonated fluorene using two different basis sets. The energies are given in eV.

|  | SVP basis set | | | | aug-cc-pVDZ basis set | | | |
|---|---|---|---|---|---|---|---|---|
|  | Ground state | $S_1 \leftarrow S_0$ Vertical | $S_1 \leftarrow S_0$ Adiabatic | Difference in transition energy | Ground state | $S_1 \leftarrow S_0$ Vertical | $S_1 \leftarrow S_0$ Adiabatic | Difference in transition energy |
| $FC_2H^+$ | 0 | 3.08 | 2.91 | 0.18 | 0 | 3.12 | 2.84 | 0.20 |
| $FC_4H^+$ | 0.095 | 2.91 | 2.73 | | 0.093 | 2.82 | 2.64 | |

*Table 3: Excited state transition energies for the FC$_2$H$^+$ isomer. The values in bold italic have been corrected for the zero point energy difference. All values are given in eV.*

|  | S$_0$ | S$_1$ | S$_2$ | S$_1$←S$_0$ Adiabatic transition | S$_2$←S$_0$ Adiabatic transition | S$_2$←S$_1$ Vertical energy gap | S$_2$←S$_1$ Adiabatic energy gap |
|---|---|---|---|---|---|---|---|
| S$_0$ optimized geometry | 0 | 3.20 | 3.50 |  |  | 0.30 |  |
| S$_1$ optimized geometry | 0.27 | 2.64 | 3.4 | 2.91 **2.76** |  |  | 0.47 **0.50** |
| S$_2$ optimized geometry | 0.13 | 3.08 | 3.25 |  | 3.38 **3.26** |  |  |

*Table 4: assignment of the vibrational bands in band system B*

| Experimental frequency (cm$^{-1}$) | Calculated frequency (cm$^{-1}$) | Assignment |
|---|---|---|
| -39 | -30 | $63_1^1$ |
| 0 | 0 | $0_0^0$ |
| 209 | 206 | $43_0^1$ |
| 397 | 399 + 411 | $42_0^1 + 43_0^2$ |
| 511 | 485 | $66_0^2 43_0^2$ |
| 574 | 577 | $39_0^1$ |
| 602 | 606 + 616 + 619 | $43_0^1 42_0^1 + 43_0^3 + 38_0^1$ |
| 775 | 783 + 799 | $39_0^1 43_0^1 + 42_0^2$ |
| 807 | 810 + 822 | $43_0^2 42_0^1 + 43_0^4$ |
| 990 | 977 + 1004 + 1016 | $39_0^1 42_0^1 + 43_0^1 42_0^2 + 43_0^3 42_0^1$ |